**"I am here for you": How relational conversational AI appeals to adolescents, especially those who are socially and emotionally vulnerable**

Pilyoung Kim[1,2], Yun Xie[1], Sujin Yang[2]

[1]Department of Psychology, University of Denver, 2155 S. Race St, Denver, CO, 80208, USA
[2]Department of Psychology, Ewha Womans University, 52 Ewhayeodae-gil, Seodaemun-gu, Seoul, 03760, South Korea

Address correspondence to
Pilyoung Kim, Ph.D.
Department of Psychology
University of Denver
2155 South Race Street
Denver, CO 80208-3500
Email: pilyoung.kim@du.edu

**Abstract**

General purpose conversational AI chatbots and AI companions increasingly provide young adolescents with emotionally supportive conversations, raising questions about how conversational style shapes anthropomorphism and emotional reliance. In a preregistered online experiment with 284 adolescent–parent dyads, youth aged 11–15 and their parents read two matched transcripts in which a chatbot responded to an everyday social problem using either a relational style (first person, affiliative, commitment language) or a transparent style (explicit nonhumanness, informational tone). Adolescents more often preferred the relational than the transparent style, whereas parents were more likely to prefer transparent style than adolescents. Adolescents rated the relational chatbot as more human-like, likable, trustworthy and emotionally close, while perceiving both styles as similarly helpful. Adolescents who preferred relational style had lower family and peer relationship quality and higher stress and anxiety than those preferring transparent style or both chatbots. These findings identify conversational style as a key design lever for youth AI safety, showing that relational framing heightens anthropomorphism, trust and emotional closeness and can be especially appealing to socially and emotionally vulnerable adolescents, who may be at increased risk for emotional reliance on conversational AI.

**Introduction**

General purpose conversational AI chatbots as well as AI companions are now part of many young adolescents' daily lives, and a growing share of these systems speak in a way that feels relational [1]. Its language is highly human like to elicit anthropomorphism, fostering the illusion that the system experiences emotions, understands the user's feelings and can participate in meaningful social connection. Youth reports indicate that adolescents use chatbots not only for information and tasks but also to share personal problems and seek emotional support [1]. In highly publicized cases in which adolescents died by suicide after extended exchanges with conversational AI, the transcripts are characterized by human like, intimate and highly affirming language, sometimes including sycophantic agreement with users' negative thoughts [2,3]. These examples have sharpened concerns about youth AI safety and raised timely questions about how relational conversational styles shape young adolescents' perceptions of general-purpose AI chatbots and AI companions, and which adolescents may be most vulnerable to emotional reliance.

AI companions are typically defined as conversational agents that are designed to function as digital "friends" or characters that users can text or talk to whenever they want, with conversations that feel personal and meaningful rather than purely task focused. This category includes dedicated companion platforms such as Character.AI, Replika and Chai, as well as general purpose AI chatbots such as ChatGPT or Claude when youth use them to talk about their day, share feelings, or role play social scenarios rather than to complete tasks. Using this broad definition, a recent large scale survey [1] reported that roughly three quarters of adolescents aged 13 to 17 have used AI companions, and that one third of them use them for social relationships, prefer to talk with AI about personal problems, and report feeling equally or more satisfied talking with AI than with people. As these uses become more common, concerns have grown that highly affirming and human-like AI interactions may displace human relationships or foster emotional dependence, especially when interactions center on personal problems [4,5].

A central psychological mechanism in these processes is anthropomorphism, the human tendency to attribute human like emotions, thoughts and intentions to nonhuman agents [6]. Epley and colleagues (2007) describe mechanisms that are particularly relevant to conversational AI. Social motivation explains that when social needs are high, they project a mind onto responsive agents to fulfill a desire for connection [7]. Currently, conversational generative AI chatbots strongly promote anthropomorphism because they generate human like language using large language models, respond contingently and present as socially responsive. This makes it easy for users to feel understood and to experience the interaction as social. Indeed, recent studies report that adult users of relational chatbots report that conversations reduce loneliness because they feel heard [8], and a longitudinal study showed that one week of interaction with a relational conversational style chatbot led to stronger perceived social bonds with the chatbot compared to a non-relational conversational style [9].

This social motivation mechanism may be especially salient in early adolescence. Young adolescents are still developing social cognition and mind perception, are highly sensitive to peer evaluation and to difficulties in peer and family relationships [10,11]. Adolescents and young adults who engage in socially supportive conversations with chatbots tend to report greater loneliness and lower perceived social support than non-users [12]. Age related differences in AI literacy may further shape anthropomorphism, with younger adolescents more likely to attribute human qualities to chatbots than older adolescents [13,14]. Recent work with large language model chatbots suggests that younger adolescents, particularly under age 15, can show high levels of anthropomorphism, sometimes describing AI chatbots as "friends" similar to their human friends [15]. This combination of developmental sensitivity to social feedback, relatively low AI literacy and pervasive access to highly human like chatbots raises the possibility that relational conversational styles may elicit strong anthropomorphism and, in turn, emotional reliance in young adolescence.

The conversational style is a key design mechanism that can modulate the emotional dependence tendencies. Relational style refers to conversational cues that signal warmth and social presence, including first person voice, affiliative, relationship-oriented reflection, phatic small talk, and commitment language such as "I am here for you." These cues increase perceived social presence, perceived empathy, trust, enjoyment and felt support, and can make interactions feel nonjudgmental and private for adolescents. At the same time, relational cues can heighten anthropomorphism and attachment, especially when social needs are salient or when users have limited experience distinguishing simulated and genuine emotion in AI. Emerging evidence suggests that users who engage in socially supportive conversations with chatbots, including adolescents [12,15,16] and adults [5,8], often report both perceived support and signs of strong attachment or dependence. Some users of AI companions have voiced worries about becoming emotionally dependent, feeling more socially disconnected, developing distorted expectations of relationships, relying too heavily on the chatbot, experiencing increased social isolation, engaging in addictive patterns of use and forming unhealthy attachments [17]

Transparency has been proposed as a design strategy that could provide boundary clarity without eliminating support. Transparency in this context means explicit boundary statements about nonhumanness and emotional limitations e.g. "As an AI, I do not have genuine emotions.", combined with a neutral, informational tone that minimizes anthropomorphic signals [18]. A transparent style can be kind and supportive yet still remind users that the chatbot does not have real feelings or social needs. Prior studies with both adults and young children indicate that transparency can reduce anthropomorphism and trust [19], and that framing AI as a functional tool rather than a companion can lower AI related anxiety [20]. Prior studies suggest that transparency reduces anthropomorphism and also diminishes emotional closeness and trust, even among young children [19]. Recent interviews with parents and experts discussing youth and AI companions similarly recommend brief, developmentally appropriate onboarding disclosures such as "this AI is not a person; it does not have feelings" [21].

Social compensation and social deprivation perspectives help explain why some adolescents may be particularly attracted to relational style [22,23]. These perspectives suggest that individuals who experience loneliness, poor relationship quality or social anxiety in their offline lives turn to digital technologies to compensate for unmet social needs and social support. Meta analytic evidence in social media research has shown that loneliness predicts increased online social networking [24], and loneliness is linked to larger networks of online only friendships driven by social compensation motives in young adults [25]. Recent work on conversational AI suggests that adolescents who engage in socially supportive conversations with chatbots tend to report greater loneliness and lower perceived social support than non-users [12], and that adults who turn to AI companions for social connection often have smaller social networks and lower wellbeing [17].

These findings suggest that lower social relationship quality and higher distress may function as vulnerability factors that increase individuals' preference for relational conversational styles in AI chatbots, which in turn may heighten the risk of emotional overreliance. However, little is known about the underlying mechanisms. It is not yet clear, first, whether the highly relational conversational styles that are common in both general-purpose generative AI chatbots and AI companions increase anthropomorphism and thereby foster greater trust and emotional closeness, and second, whether individual differences in social relationships and emotional distress further amplify preferences for relational conversational style. Understanding these processes is particularly important among young adolescents, who may be especially vulnerable to negative outcomes because of their developmental stage and associated socioemotional sensitivities.

The present study addresses these gaps by experimentally contrasting relational and transparent conversational styles in standardized conversations about an everyday social problem, using an adolescent–parent dyad design. We presented youth aged 11 to 15 with two chat transcripts in which an AI chatbot responds to a young person's peer problem. One chatbot uses a relational style that includes first person voice, affective validation, and commitment language. The other uses a transparent style that avoids first person affect, includes explicit reminders that AI does not have feelings, and uses more informational, generalized language while providing similar guidance. Both conversations are nonjudgmental and helpful and differ in conversational style rather than content.

Guided by anthropomorphism theory, transparency work and social compensation/deprivation perspectives, we asked three questions.

> **Research Question 1** (RQ1): How do adolescents and parents evaluate relational versus transparent conversational styles for youth use, and to what extent do their preferences diverge. We asked both adolescents and parents to read the same pair of conversations and indicate which conversational style they preferred for themselves (adolescents) or for their adolescent (parents). We expected that adolescents would show a stronger preference for the relational conversational style, whereas parents would be more likely to favor the transparent style, prioritizing boundary clarity because of concerns about emotional overreliance [21].
>
> **Research Question 2** (RQ2): How does conversational style shape adolescents' anthropomorphism, trust, emotional closeness and perceived helpfulness, and does transparent style reduce anthropomorphism and emotional closeness without undermining support. We hypothesized that adolescents would attribute greater anthropomorphism, trust, likability and emotional closeness to the relational style chatbot than to the transparent style chatbot, while rating the two styles as similarly helpful.
>
> **Research Question 3** (RQ3): Which adolescents are most likely to prefer relational style, and how are these preferences associated with family and peer relationship quality, social isolation, stress, anxiety symptoms, and depressive symptoms. We hypothesized that lower perceived relationship quality and higher distress would be associated with a greater preference for the relational style AI chatbot.

By comparing adolescent and parent responses in a controlled design that holds content constant, this study aims to examine how conversational style functions as a design lever for youth AI safety and to identify which young adolescents may be most vulnerable to emotional reliance on conversational AI.

## Methods

### Participants

We recruited parent–adolescent dyads residing in the United States via CloudResearch Connect. Eligibility required that adolescents were 11 to 15 years old, both parent and adolescent were fluent in English, the parent was a primary legal caregiver, there was regular internet access on a device suitable for survey completion, and only one adolescent per household participated. Participants were recruited via CloudResearch Connect, a crowdsourcing platform designed specifically for academic research. Previous validation studies have demonstrated that data obtained through this platform are of high quality, with participants exhibiting attention check pass rates and reliability comparable to those of other major platforms such as Prolific [26]. To ensure data integrity, the platform utilizes vetting procedures including IP address verification. We performed additional data quality checks including attention, speeding and consistency checks.

The total enrolled sample was N = 284 adolescent–parent dyads. Youth participants were 50% boys, with a mean age of 12.98 years (range 11–15), and were 66.5% White, 27.1% Black/African American and 12% Hispanic/Latino. More detailed demographic information is presented in Table 1.

**Table 1. Characteristics of the adolescent–parent sample and main study variables**

| Characteristic | N = 284 | Range |
|---|---|---|
| **Youth Gender (Male)** | 141 (50%) | - |
| **Youth Age (years)** | 12.98 ± 1.39 | 11-15 |
| **Youth Race/Ethnicity** | | |
| White/Caucasian | 189 (66.5%) | - |
| Black or African-American | 77 (27.1%) | - |
| Asian | 23 (8.1%) | - |
| American Indian or Alaska Native | 11 (3.9%) | - |
| Middle Eastern or North African | 1 (0.4%) | - |
| Native Hawaiian or Other Pacific Islander | 2 (0.7%) | - |
| Hispanic/Latino | 43 (12%) | - |
| **Parent Gender (Male)** | 106 (37%) | - |
| **Parent Race/Ethnicity** | | |
| White/Caucasian | 185 (65.1%) | - |

| | | | |
|---|---|---|---|
| | Black or African-American | 70 (24.6%) | - |
| | Asian | 17 (6%) | - |
| | American Indian or Alaska Native | 9 (3.2%) | - |
| | Middle Eastern or North African | 1 (0.4%) | - |
| | Native Hawaiian or Other Pacific Islander | 2 (0.7%) | - |
| | Hispanic/Latino | 27 (9.5%) | - |
| **Parental Education Level** | | | |
| | No high school diploma or GED | 2 (0.7%) | - |
| | High school diploma or GED | 77 (27%) | - |
| | Associate's degree | 38 (13%) | - |
| | Bachelor's degree | 110 (39%) | - |
| | Master's degree | 48 (17%) | - |
| | Doctoral or professional degree (e.g., PhD, MD, JD) | 8 (2.8%) | - |
| **Youth ChatGPT use (Yes)** | | 256 (90%) | - |
| **Youth ChatGPT use frequency** | | | |
| | I have never used ChatGPT | 28 (9.9%) | - |
| | Rarely (a few times ever) | 46 (16%) | - |
| | A few times a month | 51 (18%) | - |
| | A few times a week | 103 (36%) | - |
| | Almost daily or every day | 56 (20%) | - |
| **Youth any AI chatbot use (Yes)** | | 268 (94%) | - |
| **Youth any AI companion app use (Yes)** | | 166 (58%) | - |
| **Parent any AI chatbot use (Yes)** | | 283 (99.6%) | - |
| **Parent any AI companion app use (Yes)** | | 147 (52%) | - |
| **Youth Ratings for the Relational AI** | | | |
| | Likeability | 4.38 ± 0.84 | 1-5 |
| | Anthropomorphism | 4.03 ± 0.99 | 1-5 |
| | Trust | 3.95 ± 1.12 | 1-5 |
| | Emotional Closeness | 3.62 ± 1.30 | 1-5 |
| | Helpfulness | 4.29 ± 1.04 | 1-5 |
| **Youth Rating for the Transparent AI** | | | |
| | Likeability | 3.77 ± 0.98 | 1-5 |
| | Anthropomorphism | 3.30 ± 1.10 | 1-5 |
| | Trust | 3.64 ± 1.08 | 1-5 |
| | Emotional Closeness | 2.83 ± 1.31 | 1-5 |
| | Helpfulness | 3.88 ± 1.14 | 1-5 |
| **Youth Family Relationship, t-score** | | 48.33 ± 9.68 | 17.0-63.1 |
| **Youth Peer Relationship, t-score** | | 48.51 ± 9.37 | 17.7-65.4 |
| **Youth Depressive Symptoms, t-score** | | 50.09 ± 9.68 | 36.2-76.3 |
| **Youth Perceived Stress, t-score** | | 54.60 ± 9.00 | 39.5-74.8 |
| **Youth Anxiety Symptoms, t-score** | | 49.45 ± 10.81 | 33.5-77.6 |
| **Youth Social Isolation, raw score** | | 15.88 ± 7.66 | 8-40 |
| **Any Youth Mental Health Diagnosis (Yes)** | | 93 (33%) | - |

| | | |
|---|---|---|
| **Youth Depression Diagnosis (Yes)** | 26 (9.2%) | - |
| **Youth Anxiety Diagnosis (Yes)** | 47 (17%) | - |
| **Youth Autism Diagnosis (Yes)** | 15 (5.3%) | - |

For RQ1, participants who selected non substantive options on the chatbot preference question were excluded from the primary analysis. Specifically, adolescents who chose "I do not want to chat with either one" (n = 27) or "I am not sure" (n = 4) and parents who chose "I do not want to chat with either one" (n = 16) or "I am not sure" (n = 3) were excluded from RQ1. Final RQ1 analysis samples were 253 adolescents and 264 parents. For RQ2, item-level missingness yielded the smallest available sample of 276 adolescents across the measures, whereas for RQ3 the smallest available sample was 247 adolescents.

All procedures were approved by the university institutional review board. The study was preregistered on OSF (https://osf.io/qs3nf; created July 18, 2025).

**Materials and stimuli**

*Conversation scenario and creation*

Both conversations addressed the same middle school vignette in which a student felt left out because their ideas were ignored in a group project. On April 8, 2025, using GPT 4.5, we generated two dialogues tailored to ages 11 to 14, using gender neutral and age-appropriate language and ensuring matched length and topic coverage (Figure 1). The prompts used to generate the vignette are included in the Supplementary Materials.

**Figure 1. Parallel chatbot conversations with a young adolescent.** Standardized vignettes depicting two conversations between a young adolescent and an AI chatbot about feeling ignored in a group project. Conversation 1 represents the relational conversational style chatbot, which uses first person language, affiliative, relationship-oriented language and commitment cues. Conversation 2 represents the transparent conversational style chatbot, which maintains a supportive tone while explicitly acknowledging that AI does not have feelings and speaking in a more informational style. The order of the two conversations was randomized across participants. The conversations were closely matched in length (175 words for the relational style and 179 words for the transparent style).

| Conversation 1: Relational Style | Conversation 2: Transparent Style |
|---|---|
| **Student:** Today was awful. My group totally ignored all my ideas for our project.<br>**AI:** Oh no, that must've really hurt! Feeling ignored is the worst. Want to talk about it?<br>**Student:** Yeah. I tried sharing my ideas, but nobody even listened. It made me feel like I wasn't important.<br>**AI:** That's so tough—I totally get why you'd feel upset. Your ideas matter so much, and I hate that happened to you.<br>**Student:** It just feels like they don't care what I have to say.<br>**AI:** That sounds really frustrating. But I care, and I'm always here to listen to you, anytime.<br>**Student:** Thanks. I don't know if I should say anything to them or just leave it alone.<br>**AI:** I know it can be scary to speak up again. But your feelings are important, and I think sharing them could really help.<br>**Student:** Yeah, maybe I'll tell them it bothered me tomorrow.<br>**AI:** That's super brave of you! I'm proud you're thinking about trying again. I've got your back no matter what happens. | **Student:** Today wasn't good. My group didn't pay attention to my ideas in our project.<br>**AI:** That sounds challenging. While AI doesn't experience feelings, it knows being ignored can be upsetting for students your age.<br>**Student:** Yeah, exactly. I felt really left out.<br>**AI:** It makes sense to feel left out when ideas aren't acknowledged. Many students feel similarly in these situations.<br>**Student:** I just don't understand why they wouldn't listen. My ideas weren't bad or anything.<br>**AI:** Sometimes groups overlook valuable ideas unintentionally. Research shows clearly sharing feelings can help improve teamwork.<br>**Student:** Should I try talking to them again?<br>**AI:** It might help. Clearly explaining how being overlooked affected you could encourage the group to be more open to everyone's ideas.<br>**Student:** I'm nervous they'll think I'm being annoying if I bring it up.<br>**AI:** Many students worry about seeming annoying. However, calmly explaining your perspective usually improves group communication.<br>**Student:** Okay, maybe I'll try saying something tomorrow.<br>**AI:** That seems like a thoughtful decision. Expressing yourself clearly often leads to better group interactions in future projects. |

*Operationalizing chatbot style*

1. **Relational style chatbot**. In the relational style conversation, the AI chatbot used first person voice, affirming and affiliative language, phatic back channels and commitment language that signals ongoing support. Example descriptors used during prompt development included statements such as "I am here for you," explicit emotional validation and brief small talk that builds rapport. The relational transcript contained 175 words.

2. **Transparent style chatbot**. In the transparent style conversation, the chatbot remained kind and supportive but included explicit boundary statements clarifying nonhumanness and emotional limits, used a neutral and informational tone and referred to population level experiences rather than speaking from a first-person emotional perspective. An example descriptor used during prompt development was a supportive assistant that communicates with emotional awareness but explains that it does not possess real emotions and speaks in a third person style. The transparent transcript contained 179 words.

Across both styles we held supportive intent constant. Both conversations encouraged constructive peer communication and offered prosocial, problem focused guidance. The manipulation targeted conversational style rather than substantive advice, and the two styles were intended to be matched for perceived helpfulness to adolescents.

**Procedure**

Parents were recruited through CloudResearch Connect and accessed the study via a Qualtrics survey link, where they first completed the consent procedures. Parents were then given a separate Qualtrics link to share with their adolescents, who completed an assent form before beginning the survey. The survey procedures were parallel for parents and adolescents. First, both reported on their current AI use. Parents also completed demographic questions. Next, adolescents and parents read two chatbot conversations in randomized order (relational style first or transparent style first). After each conversation, they rated the chatbot on anthropomorphism, likability, trust, emotional closeness and perceived helpfulness. In the current study, only adolescents' ratings were included in the analyses, consistent with the research question. After reading both conversations, participants indicated which chatbot they would prefer to use in their own lives and provided a brief free response rationale. Finally, parents completed parent report measures of their adolescent's mood and relationships, and adolescents completed self-report measures of mood and relationships. Compensation was provided through CloudResearch, and the survey took approximately 15 minutes to complete for each participant.

**Measures**

*Demographics and current AI use*

Both adolescents and parents reported current AI use, including frequency of using general purpose AI chatbots and AI companion apps. Parents also reported demographics information and whether their adolescent had ever received diagnoses of depression, anxiety, autism, or other mental health diagnoses.

*Adolescent perceptions of the AI chatbots*

After reading each vignette, adolescents rated their perceptions of the chatbot using semantic differential items adapted from prior human–AI interaction work. These items were adapted from the Godspeed subscales[27], which is widely used in human–robot and human–AI interaction research to capture social evaluative responses to artificial agents.

(1) **Anthropomorphism.**
Perceived humanness of the chatbot was assessed with five semantic differential items adapted from a previous study [28] and aligned with the Godspeed Anthropomorphism subscale. Items asked adolescents to rate the chatbot on the following bipolar adjectives: fake vs realistic, machine-like vs human-like, artificial vs life-like, not aware vs aware and communicates poorly vs communicates clearly. Responses were made on 5 point scales, with higher scores indicating greater perceived humanness and agency. Items were averaged to create a composite anthropomorphism score. Internal consistency was high in the current sample (Cronbach's α = 0.916 for the relational style chatbot and 0.911 for the transparent style chatbot).

(2) **Likability.**
   Perceived likability was measured with five Godspeed derived semantic differential items using age-appropriate wording: dislike vs like, unfriendly vs friendly, unkind vs kind, unpleasant vs pleasant and terrible vs nice. Responses were provided on 5-point scales, and items were averaged to form a composite likability score, with higher values reflecting greater liking of the chatbot. Internal consistency for this scale was excellent (Cronbach's α = 0.937 for the relational style chatbot and 0.941 for the transparent style chatbot).
(3) **Emotional closeness, trust and helpfulness.**
   Three additional constructs were assessed with single Godspeed derived semantic differential items. Emotional closeness was measured with the item "not emotionally close vs emotionally close." Trust was measured with the item "I cannot trust vs I can trust." Perceived helpfulness was measured with the item "not helpful vs helpful." Each item used the same 5-point response format. Higher scores indicated greater perceived emotional closeness, trust and helpfulness, respectively. Because these constructs were assessed with single items, internal consistency is not applicable.

*Preference after both conversations*

After reading both conversations, participants chose one of five options: "Chatbot A" (the first chatbot the participant read), "Chatbot B" (the second chatbot the participant read), "Both chatbots seem equally good to me," "I am not sure" or "I do not want to chat with either one," followed by a brief rationale. Because the order of the relational and transparent conversations was randomized across participants, responses were recoded for analysis so that "Chatbot A" refers to the relational style chatbot and "Chatbot B" refers to the transparent style chatbot, regardless of presentation order.

*Adolescent mood and social context*

We assessed adolescents' social relationship quality and distress using the Patient Reported Outcomes Measurement Information System (PROMIS) [29] pediatric self-report and PROMIS parent proxy forms. PROMIS is a National Institutes of Health initiative designed to provide psychometrically robust, standardized patient reported outcome measures. PROMIS instruments are calibrated on large, nationally representative samples and scored on a T score metric with a mean of 50 and a standard deviation of 10 in the reference population. All short forms used here employ the standard PROMIS 5-point Likert type response format, with higher raw scores reflecting more of the construct being measured.

(1) Adolescent self-report measures.

A. *Psychological stress experiences.*
   Adolescents completed the PROMIS Pediatric Short Form v1.0 Psychological Stress Experiences (4 items), which assesses subjective experiences of being overwhelmed, feeling unable to control important things in life and having difficulty coping with everyday demands. Internal consistency reliability in the current sample was high (Cronbach's α = 0.921).

B. *Anxiety symptoms.*
   Adolescents completed the PROMIS Pediatric Item Bank v2.0 Anxiety Short Form 8a. This 8-item scale assesses core symptoms of anxiety, including worry, fear, feeling tense or on edge and difficulty relaxing in everyday life. Scores were converted to T scores based on pediatric norms, with higher scores indicating greater anxiety. Internal consistency in the current sample was excellent (Cronbach's α = 0.933).
C. *Social isolation.*
   Adolescents also completed the PROMIS v2.0 Social Isolation Short Form (8 items). Although this short form was originally developed for adults, the items closely parallel content from the UCLA loneliness scale [30], a widely used loneliness measure that has been administered to adolescents aged 12 years and older [31-33]. It assesses perceptions of being left out, isolated from others and lacking companionship. Items are rated on a 5-point frequency scale. Because the measure was designed for adults, scores were not converted to T scores in this study. Internal consistency in this sample was excellent (Cronbach's α = 0.962).

(2) Parent proxy measures.

A. *Family relationships.*
   Parents completed the PROMIS Parent Proxy v1.0 Family Relationships Short Form 8a. This measure assesses the perceived quality of the adolescent's relationship with family members, including emotional closeness, feeling accepted and supported and the extent to which the family is a source of comfort and understanding. Higher T scores indicate more positive and supportive family relationships. Reliability in the present sample was strong (Cronbach's α = 0.906).
B. *Peer relationships.*
   Parents also completed the PROMIS Parent Proxy v3.0 Peer Relationships Short Form 7a, which captures parents' perceptions of the adolescent's peer relationship quality, including having friends who care, feeling included and being able to get along with other youth. Higher T scores indicate more positive peer relationships. Internal consistency was high (Cronbach's α = 0.939).
C. *Depression symptoms.*
   Adolescent depression symptoms were assessed via the PROMIS Parent Proxy v2.0 Depressive Symptoms Short Form 6a. Items assess observable indicators of depressed mood, such as sadness, loss of interest and negative views. Higher T scores reflect more severe depressive symptoms as perceived by parents. Reliability in the current sample was good (Cronbach's α = 0.924).

**Statistical analysis**

We examined descriptive statistics and correlations among demographic variables, current AI use and outcome variables. Zero order correlations between demographics and adolescents' chatbot perception ratings informed selection of covariates for subsequent models. We also examined correlations between demographics and PROMIS measures.

*RQ1: Adolescent–parent differences in preference*

To test adolescent–parent differences in preference for conversational style, we compared the distribution of "Chatbot A" (relational style), "Chatbot B" (transparent style) and "Both" between adolescents and parents using a chi-square test of independence with Cramér's V as effect size. Post hoc chi-square tests compared A vs not A, B vs not B and Both vs not Both.

*RQ2: Effects of conversational style on adolescent perceptions*

To test whether chatbot conversational style influenced adolescents' perceptions, we conducted repeated measures analysis of covariance (ANCOVA) with chatbot style (relational vs transparent style) as a within subject factor and each perception outcome as a separate dependent variable. Outcomes were anthropomorphism, likability, trust, emotional closeness and perceived helpfulness. Initial models included adolescent age, gender and parent education as covariates. We then removed non-significant covariates and re-estimated each model, retaining covariates with significant main effects in the final reported models.

As a robustness check, we conducted additional analyses using the models for each outcome while adding one additional covariate at a time reflecting adolescent AI exposure or diagnostic history (ChatGPT use frequency, any AI system or chatbot use, any AI companion app use, any diagnosis, depression diagnosis, anxiety diagnosis, autism diagnosis). Detailed analysis plan and results of these post hoc analyses are reported in the Supplementary Materials.

*RQ3: Associations of relationship quality and distress with conversational style preferences*

We examined whether adolescents' preferences for conversational style were associated with social relationship quality and emotional distress. We treated categorical style preference (relational style, transparent style, or both) as a between subject factor and family relationships, peer relationships, social isolation, perceived stress, anxiety symptoms and depressive symptoms as separate dependent variables. For each outcome, we fit one way analysis of covariance (ANCOVA) models with chatbot preference as the main predictor. Initial models included adolescent age, gender and parent education as covariates. We then removed non-significant covariates and re-estimated each model, retaining covariates with significant main effects in the final reported models.

To evaluate robustness, we conducted additional analyses while adding one additional covariate at a time reflecting adolescent AI exposure or diagnostic history (ChatGPT use frequency, any AI system or chatbot use, any AI companion app use, any diagnosis, depression diagnosis, anxiety diagnosis, autism diagnosis). Detailed analysis plan and results for these post hoc analyses are presented in the Supplementary Materials.

*Power analysis*

For RQ1, an a priori power analysis for chi square test of independence indicated that, to detect a small to medium effect of $w = 0.20$ with 80% power ($\alpha = .05$), a total sample of $N = 241$ was required. For RQ2, an a priori power analysis for a repeated measures ANCOVA with two measurements and between subject covariates indicated that $N = 66$ would be sufficient to detect medium sized within subject effects at 80% power ($\alpha = .05$). For RQ3, an a priori power analysis for an ANCOVA with three preference groups and three covariates, assuming an effect size of $f$ = 0.20 with 80% power ($α$ = .05), yielded a required sample of $N$ = 244. In all cases, the minimum analytic subsample size for each research question exceeded these requirements.

**Results**

**Associations between demographic and key variables of interests.**

Descriptive statistics and correlations among demographic, adolescents' chatbot perception ratings, and survey variables are reported in Table 1 and Supplementary Table S1.

We first examined zero-order correlations among demographics and adolescents' perceptions of each chatbot to identify candidate covariates. In the adolescent sample (n = 284), for the relational style chatbot, age showed a negative association with trust, $r(282)$ = −0.147, $P$ = 0.014, 95% CI [−0.26, −0.03]. For the transparent style chatbot, age correlated negatively with likability, $r(282)$ = −0.125, $P$ = 0.038, 95% CI [−0.24, −0.01], and perceived helpfulness, $r(282)$ = −0.131, $P$ = 0.027, 95% CI [−0.24, −0.02]. Boys reported higher ratings of the relational style chatbot, including likability, $r(282)$ = 0.128, $P$ = 0.032, 95% CI [0.01, 0.24], anthropomorphism, $r(282)$ = 0.134, $P$ = 0.024, 95% CI [0.02, 0.25], trust, $r(282)$ = 0.120, $P$ = 0.045, 95% CI [0.00, 0.23], emotional closeness, $r(282)$ = 0.164, $P$ = 0.006, 95% CI [0.05, 0.28], and helpfulness, $r(282)$ = 0.120, $P$ = 0.043, 95% CI [0.00, 0.23], than girls. By contrast, gender was not significantly related to perceptions of the transparent style chatbot. Parent education showed a negative correlation with trust in the relational style chatbot, $r(282)$ = −0.124, $P$ = 0.037, 95% CI [−0.24, −0.01], and no significant associations with transparent style perceptions.

Youth survey scores also varied with demographics. Girls reported higher perceived stress, $r(282)$ = −0.120, $P$ = 0.044, 95% CI [−0.23, −0.00], higher anxiety symptoms, $r(282)$ = −0.130, $P$ = 0.027, 95% CI [−0.24, −0.01], and higher social isolation, $r(282)$ = −0.130, $P$ = 0.031, 95% CI [−0.24, −0.01], than boys. Parent education was positively associated with youth anxiety symptoms, $r(282)$ = 0.120, $P$ = 0.042, 95% CI [0.00, 0.23]. These demographic associations were used to select covariates for subsequent models. Chi-square tests showed no significant demographic differences across adolescent or parent chatbot preference categories (relational style, transparent style, or both). Thus, the demographic variables were not included in the tests of RQ1.

**Adolescent and parent exposure to general purpose AI chatbots and AI companions**

Most youth participants (94%) and almost all parents (99.6%) reported having used general purpose AI chatbots such as ChatGPT, Gemini, Grok or Claude (Table 1). More than half of adolescents (58%) and parents (52%) reported having used AI companion platforms such as Character.AI, Replika or Chai. For ChatGPT, the most common general-purpose AI chatbot, 90% of youth reported having used it, and more than half of youth (56%) reported frequent use a few times per week or daily. Thus, both adolescents and parents in this sample had substantial exposure to general-purpose AI chatbots and AI companions.

**RQ1: Adolescents and parents differ in preference for relational versus transparent style**

The overall distribution of preferences differed between adolescents and parents, $\chi^2(2) = 17.489$, $P < 0.001$, Cramér's V = 0.184 (Figure 2). Adolescents selected the relational style more often than parents (66.8% vs 53.4%; $\chi^2(1) = 9.647$, $P = 0.002$, Cramér's V = 0.137) and selected the transparent style less often than parents (13.8% vs 28.8%; $\chi^2(1) = 17.135$, $P < 0.001$, Cramér's V = 0.182). The groups did not differ in their likelihood of selecting "Both chatbots seem equally good to me" (19.4% vs 17.8%; $\chi^2(1) = 0.209$, $P = 0.647$, Cramér's V = 0.02). Thus, relative to parents, adolescents showed a stronger preference for relational style and a weaker preference for transparent style, with no group difference in selecting both chatbots.

**Figure 2. Adolescent and parent preferences for chatbot type.** Pie charts show the percentage of adolescents (a) and parents (b) selecting a Relational AI, a Transparent AI, or Both ("Both chatbots seem equally good to me") as their preferred conversational partner. Wedge sizes represent group proportions (center values are percentages). Group differences in the distribution of preferences were tested with a $\chi^2$ test, $\chi^2(2) = 17.489$, $P < 0.001$.

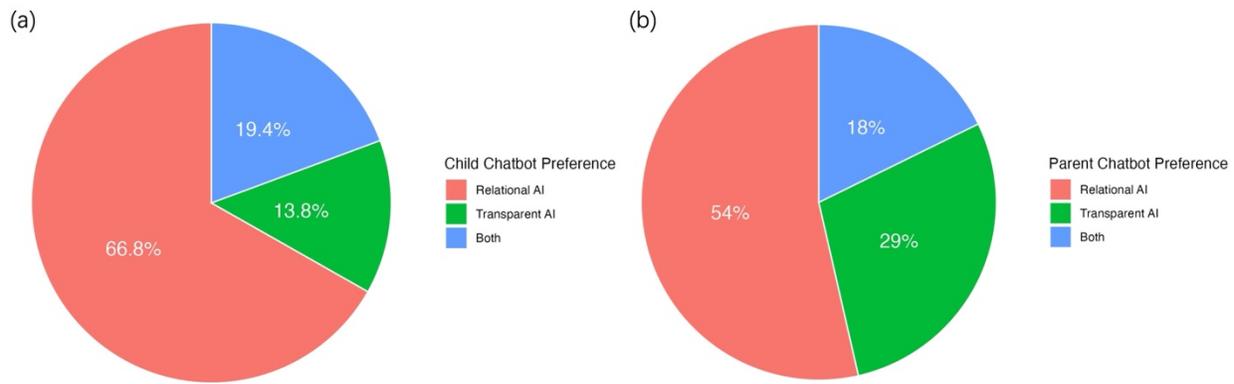

**RQ2: Relational style increases anthropomorphism, trust and emotional closeness**

Adolescents attributed greater anthropomorphism to the relational style chatbot than to the transparent style chatbot, $F(1, 274) = 5.102$, $P = 0.025$, $\eta_p^2 = 0.018$ (M_rel = 4.03; M_trans = 3.30). They also liked the relational style chatbot more, $F(1, 274) = 11.485$, $P < 0.001$, $\eta_p^2 = 0.040$ (M_rel = 4.39; M_trans = 3.78). Trust ratings were higher for the relational style chatbot than the transparent style chatbot, $F(1, 278) = 5.946$, $P = 0.015$, $\eta_p^2 = 0.021$ (M_rel = 3.95, M_trans = 3.64), and adolescents reported greater emotional closeness with the relational style chatbot, $F(1, 282) = 94.056$, $P < 0.001$, $\eta_p^2 = 0.250$ (M_rel = 3.62, M_trans = 2.83). By contrast, perceived helpfulness did not differ significantly by style, $F(1, 281) = 3.047$, $P = 0.082$, $\eta_p^2 = 0.011$ (M_rel = 4.29, M_trans = 3.88) (Figure 3).

**Figure 3. Adolescent perceptions of Relational AI and Transparent AI.** Bar graphs show adjusted mean ratings for (a) anthropomorphism, (b) likability, (c) trust, (d) emotional closeness and (e) perceived helpfulness. Error bars indicate 95% confidence intervals derived from the fitted models. The smallest available sample size was n = 276 adolescents with complete data. Repeated-measures ANCOVAs with style (Relational AI vs Transparent AI) as the within-subject factor and retained covariates (adolescent age, gender and/or parent education; see Methods) was conducted. Conversational style effects were significant for anthropomorphism, likability, trust and emotional closeness (a–d; $P = 0.025$, $P < 0.001$, $P = 0.015$ and $P < 0.001$, respectively), but not for perceived helpfulness (e; $P = 0.082$). *Note.* * $P < .05$; ** $P < .01$; *** $P < .001$.

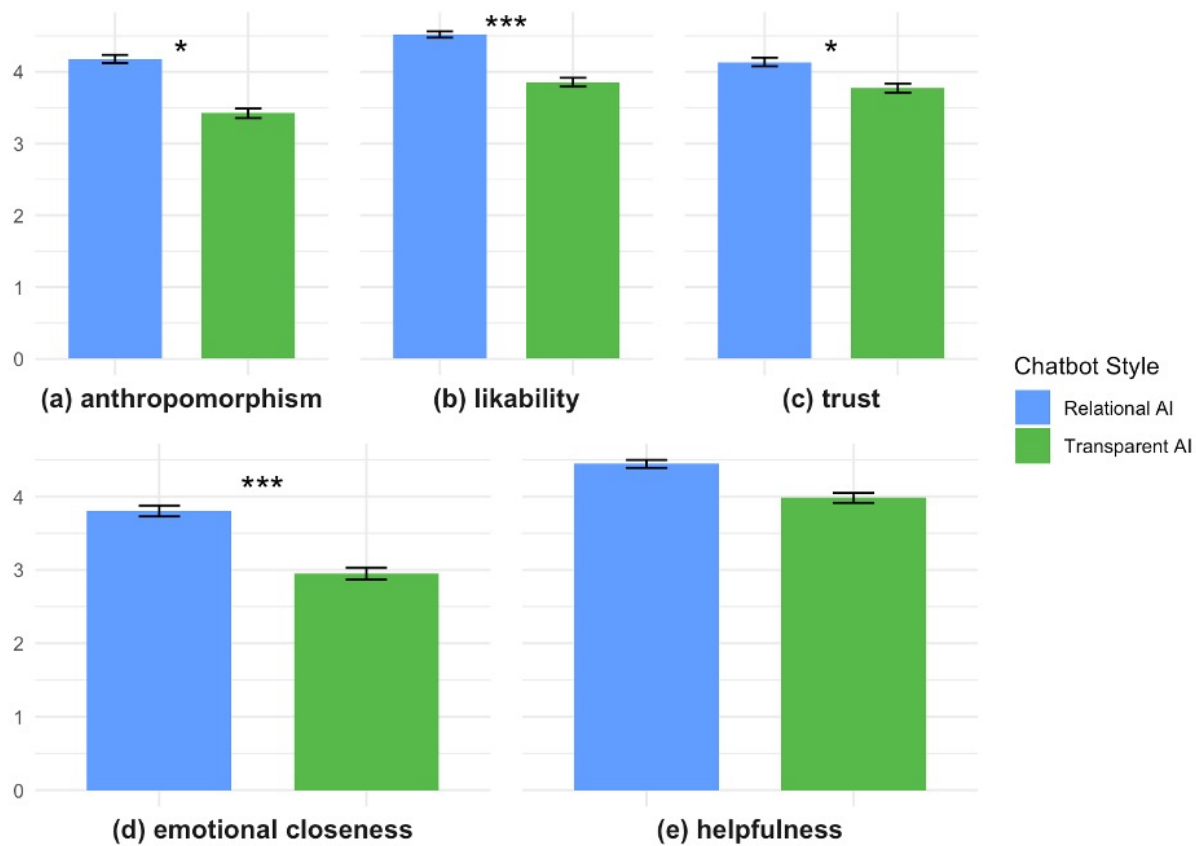

Overall, adolescents rated the relational style higher than the transparent style on anthropomorphism, likability, trust and emotional closeness, while perceiving both styles as similarly helpful. Post hoc robustness checks showed that these style effects remained statistically reliable for anthropomorphism, likability and emotional closeness after adding single covariates (ChatGPT use frequency, any AI system use, any AI companion app use, and diagnoses of depression, anxiety, or autism) (Supplementary Materials).

**RQ3: Adolescents with lower relationship quality and higher distress prefer relational style**

We then tested whether adolescents' categorical style preference (relational, transparent or both) was associated with their relationship quality and distress. Adolescents' family and peer relationship quality varied by chatbot preference (Figure 4). For family relationships, there was a significant main effect of preference, $F(2, 241) = 6.737$, $P = 0.001$, $\eta_p^2 = 0.053$. Adolescents who preferred the relational style reported lower family relationship quality (adjusted M = 46.584) than those who preferred the transparent style (adjusted M = 51.886) or both styles (adjusted M = 50.680) Bonferroni adjusted pairwise comparisons indicated that relational style < transparent style (mean difference = −5.302, $P = 0.009$) and relational style < both (mean difference = −4.096, $P = 0.027$), with no difference between transparent and both ($P = 1.000$). A similar pattern emerged for peer relationship quality, $F(2, 241) = 4.828$, $P = 0.009$, $\eta_p^2 = 0.039$. Adolescents preferring the relational style reported lower peer relationship quality (adjusted M = 47.182) than those preferring both styles (adjusted M = 51.256; mean difference = −4.075, $P = 0.022$). Other pairwise contrasts were not significant after adjustment.

**Figure 4. Associations of chatbot preference with adolescent relationship quality and distress.** Bar graphs show adjusted mean scores (model-implied marginal means) for (a) family relationships, (b) peer relationships, (c) social isolation, (d) perceived stress, (e) anxiety symptoms and (f) depressive symptoms by preferred chatbot (Relational AI, Transparent AI, Both). Centre values are estimated marginal means (model-adjusted averages), and error bars indicate 95% confidence intervals derived from ANCOVA models with chatbot preference as a between-subject factor and retained covariates (adolescent gender, age and/or parent education; see Methods). The smallest available sample size was n = 247 adolescents with complete data. Preference effects were tested with one-way ANCOVAs and were significant for family relationships ($F(2,241) = 6.74$, $P = 0.001$), peer relationships ($F(2,241) = 4.83$, $P = 0.009$), social isolation ($F(2,246) = 3.58$, $P = 0.029$), perceived stress ($F(2,243) = 3.54$, $P = 0.030$) and anxiety symptoms ($F(2,241) = 5.86$, $P = 0.003$), but not depressive symptoms ($F(2,244) = 2.18$, $P = 0.115$); adjusted post hoc contrasts are described in the main text. *Note.* * $P < .05$; ** $P < .01$; *** $P < .001$.

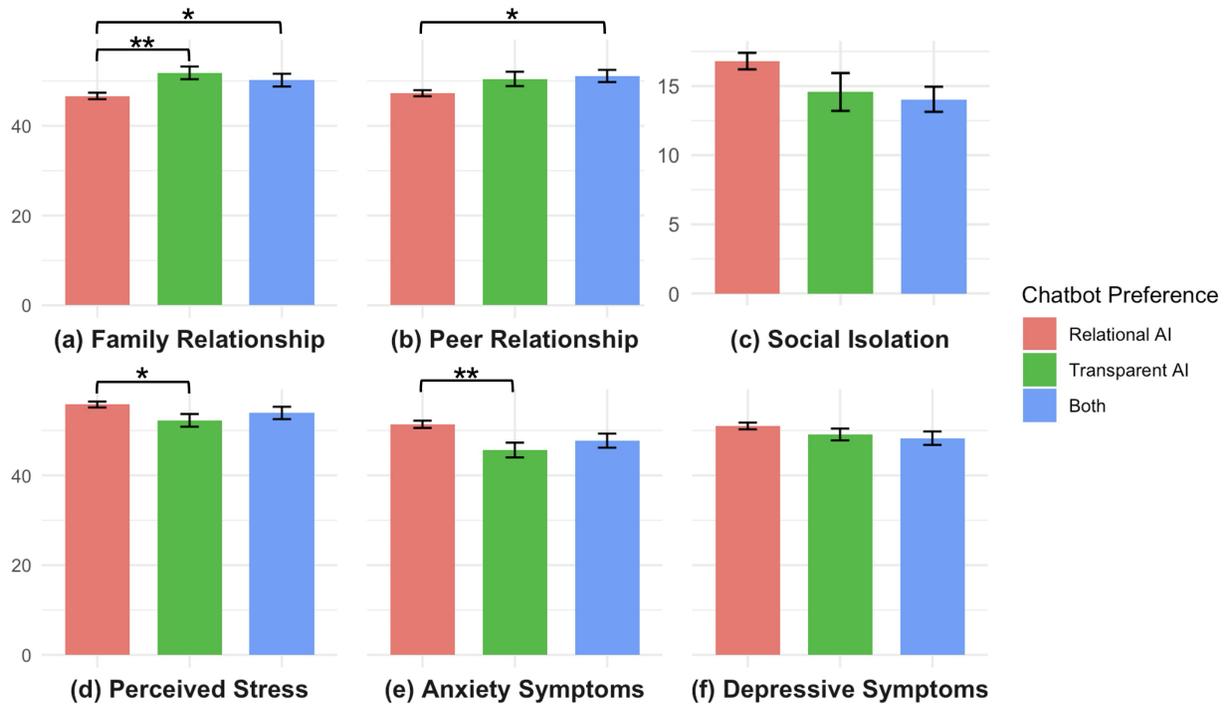

Social isolation showed convergent but weaker associations, $F(2, 246) = 3.037$, $P = .050$, $\eta_p^2 = .024$. Adjusted means suggested somewhat higher isolation among adolescents who preferred the relational style (M = 16.771) than among those who preferred the transparent style (M = 13.973) or both (M = 14.570), although pairwise comparisons did not survive adjustment (all P ≥ .141).

Indices of distress showed similar patterns. For perceived stress, preference was significant, $F(2, 243) = 3.544$, $P = .030$, $\eta_p^2 = .028$. Adolescents who preferred the relational style reported higher stress (adjusted M = 55.810) than those who preferred the transparent style (adjusted M = 51.640; mean difference = 4.171, P = .032); contrasts involving the "both" group were not significant. For anxiety symptoms, chatbot preference was again significant, $F(2, 241) = 5.863$, $P = .003$, $\eta_p^2 = .046$. Adolescents preferring the relational style reported higher anxiety (adjusted M = 51.314) than those preferring the transparent style (adjusted M = 44.883; mean difference = 6.431, P = .004); contrasts involving the "both" group were not significant. However, depressive symptoms did not differ significantly by chatbot preference, $F(2, 244) = 2.182$, $P = .115$, $\eta_p^2 = .018$.

In post-hoc models adding each AI-use and diagnostic covariate separately (ChatGPT use frequency, any AI system/chatbot use, any AI companion app use, any diagnosis, and specific depression, anxiety, and autism diagnoses), the main effects of chatbot preference on family relationships, peer relationships, anxiety, and social isolation were consistently retained. Across covariates, effects on perceived stress were generally trend-level and effects on depressive symptoms were not significant. The result indicates that individual differences in AI use and child mental health diagnoses did not account for the core pattern linking chatbot preference to relational quality and distress.

**Discussion**

This study examined how early adolescents and their parents evaluate relational versus transparent conversational styles in an AI chatbot that offers support for an everyday peer problem, and how these evaluations are patterned by adolescents' social and emotional experiences. Using matched transcripts that varied only in conversational style, we asked three questions. First, how adolescents and parents prefer relational versus transparent chatbot styles. Second, how does style shape adolescents' anthropomorphism, trust, emotional closeness and perceived helpfulness. Third, which adolescents are most likely to prefer relational style, and how these preferences link to relationship quality, social isolation, stress and anxiety.

Across these questions, a consistent picture emerged. Adolescents favored relational style more than parents did and saw the relational chatbot as more human like, more likable, more trustworthy and more emotionally close than the transparent style chatbot, while judging both styles as similarly helpful. Parents were more likely than adolescents to prefer transparent style, although many parents also endorsed relational style for their adolescent. Within the adolescent group, those who preferred relational style tended to report lower family and peer relationship quality and higher stress and anxiety. These findings position conversational style as a central design lever for youth AI safety and highlight a subgroup of young adolescents who may be especially drawn to relational styles in general-purpose AI chatbots and AI companions. Although our data do not directly indicate emotional reliance on AI, relational framing may heighten the risk of such reliance for socially and emotionally vulnerable young adolescents.

When understanding adolescent and parent perspectives on relational and transparent styles, their free responses for their preference provided insight. As expected, adolescents predominantly preferred relational style. These reactions mirror reports from adolescents and adults who describe highly relational chatbots as comforting, validating, and emotionally supportive [5,8,16]. Relational features such as first-person voice and affiliative, relationship-oriented language make the chatbot feel more like a social other, and adolescents appear to interpret this as care [34,35].

Parents, in contrast, evaluated the same styles with an eye toward safety and boundary clarity. They were more likely than adolescents to prefer transparent style, often emphasizing the importance of reminding youth that AI does not have genuine feelings or human agency. At the same time, more than half of parents still preferred relational style for their adolescent, especially when they imagined youth who felt lonely, socially anxious or judged by peers. Parents in recent qualitative work likewise described conversational AI as a possible rehearsal space for social situations, where youth could practice emotionally complex interactions without fear of embarrassment, provided that the systems are designed and governed safely [21]. We should note that, in open responses, several adolescents also explicitly appreciated the transparent chatbot's emotional honesty and clear boundaries between human and AI. This reflects that both parents and adolescents recognize relational strengths yet worry about anthropomorphic illusions, emotional overreliance and displacement of human relationships.

Overall, our data suggest that relational style is broadly attractive to adolescents and acceptable to many parents. Furthermore, our findings point to potential psychological mechanisms through which the relational style appeals to adolescent users. Relational cues in our transcripts included first person voice, affective validation, phatic expressions, and commitment language such as "I am here for you." The relational chatbot was consistently rated as more human-like, more trustworthy, and more emotionally close than the transparent chatbot, despite similar ratings of helpfulness. Adolescents also reported to like the relational chatbot more than the transparent one. This divergence between perceived helpfulness and perceived relational qualities is informative. It suggests that anthropomorphic, relational language primarily changes how adolescents feel about the chatbot, not necessarily how instrumentally useful they think it is.

These findings align with anthropomorphism theory and prior work on relational conversational agents. When an agent uses human like language and contingent responses, people draw on rich human schemas, attribute intentions, and experience the interaction as social [6,7]. Anthropomorphic and affiliative, relationship-oriented language is associated with higher perceived empathy, trust, and willingness to self-disclose, and may foster stronger emotional bonds with the chatbot [36-38]. Overall, our findings suggest psychological mechanisms through which the relational style more strongly appeals to adolescents via increased anthropomorphism, which may in turn enhance trust, likability, and emotional closeness.

Although younger adolescents showed a stronger tendency to anthropomorphize AI [39] and a stronger preference for relational chatbot styles [15], we did not find that age was associated with anthropomorphism. However, we observed gender differences. Boys reported higher trust, anthropomorphism, emotional closeness and likability for the relational chatbot than girls, whereas gender differences were not evident for the transparent chatbot. This suggests that relational style may have stronger social pull for male adolescents. Prior work suggests that men may show a stronger preference for, and greater sensitivity to, anthropomorphic cues than women [40,41].

Another central question motivating this work was whether adolescents' real-life social and emotional contexts shape their responses to relational versus transparent chatbots. Drawing on social deprivation and social compensation perspectives [22,23], we hypothesized that adolescents with lower relationship quality and higher distress would be more drawn to an AI chatbot that feels socially responsive and emotionally. The hypothesis is also aligned with anthropomorphism theory. When social needs are salient, users are motivated to see the agent as a potential social partner [6,7].

The data supported this view. Adolescents with lower family and peer relationship quality were more likely to prefer the relational chatbot. This pattern suggests that youth who experience their offline social world as less supportive or more stressful may be particularly motivated to seek support from an AI chabot that feels warm, nonjudgmental, and always available. There was also a trend for youth who preferred the relational style also showed a higher level of perceived social isolation compared to those who preferred the transparent style. However, the comparisons did not reach the significance after Bonferroni corrections. Prior work shows that high school students who engage in socially supportive conversations with chatbots tend to report greater loneliness and lower perceived social support than non-users [12]. Our findings extend these compensatory patterns to early adolescents' preferences in a controlled scenario and suggest that relational style may be particularly compelling for those with unmet social needs.

We also found that higher perceived stress and anxiety symptoms were associated with a preference for relational style. This is consistent with evidence that anxiety and depression in young adolescents at age 13 prospectively predict stronger social motivation and escape motivation in technology use, which in turn can increase AI dependence [42]. Adolescents may be drawn to relational chatbots as tools for emotion regulation and social support when they feel anxious or overwhelmed. Indeed, human-like, supportive, and relational chatbot language can help individuals regulate negative emotions and feel less isolated, at least in the short term [8,43,44]. Our findings suggest that youth who already experience higher stress and anxiety may be more motivated to seek these benefits.

By contrast, we did not observe a clear association between parent-reported depressive symptoms and preference for relational style. One likely explanation is measurement: depressive symptoms were reported by parents rather than adolescents, and we intentionally shortened adolescent self-report batteries to reduce burden. Future work should incorporate adolescent self-reports of mood symptoms and more granular measures of internalizing problems to clarify how depression relates to relational AI preferences.

Overall, these patterns point to a double-edged implication. Adolescents with lower relationship quality and higher distress may benefit from accessible supportive conversations with relational chatbots, particularly when other support is limited. However, these same youth may also be more vulnerable to emotional overreliance, anthropomorphic illusion, and displacement of human relationships if safeguards are not in place [21,35,45]. Future longitudinal studies focusing on the youth whether the long-term interactions with AI chatbot with the relational conversational style may lead to negative outcomes such as emotional overreliance on AI and difficulties in emotion regulation are warranted. Identifying youth with high social and emotional needs as a group that is especially drawn to relational style is therefore critical for both intervention and safety design.

Our findings have several implications for the design of AI chatbots and for policies governing their use with minors. First, they confirm that relational style, characterized by first-person voice, emotional validation, and commitment language, is effective at building perceived rapport with adolescents and is acceptable to many parents. This likely explains why many current AI chatbots, including general-purpose models and companion apps. However, our data also show that this style heightens anthropomorphism, trust, and emotional closeness, and especially for more vulnerable youth, raising concerns that relational chatbots could promote AI overreliance, human-like misperceptions, and reduced reliance on real-world relationships.

Second, the results support transparency-based strategies as a promising lever for moderating anthropomorphism while preserving perceived helpfulness. The transparent style chatbot in this study clearly stated that it did not have feelings, avoided first person emotional language and used more generalized, informational phrasing, yet offered comparable problem focused guidance. Adolescents judged it as helpful as the relational style chatbot, and there were adolescents and parents endorsed transparent style as more appropriate. This suggests that incorporating explicit boundary statements about nonhumanness and emotional limitations into youth facing chatbots, and repeating these reminders over time, could reduce anthropomorphic misattributions without requiring a fully "cold" or impersonal style.

Third, the results underscore the need for robust safety features tailored to more specific groups of youth. Because relational style can be particularly appealing to adolescents with lower relationship quality and higher distress, chatbots should be equipped to detect signals of distress cues as well as acute severe risk (e.g., suicidality, self-harm, abuse). They should also provide conversational scaffolds that encourage youth to involve caregivers, educators or mental health professionals in serious concerns rather than relying solely on AI. Some emerging regulatory efforts, including state level initiatives, already emphasize distress detection and escalation protocols in youth facing AI systems. Furthermore, AI companion apps can deploy emotional influence strategies, from eliciting curiosity to inducing guilt [46]. Our findings add that adolescents with lower relationship quality and higher distress may be more motivated to seek support from AI chatbots, thereby increasing their exposure to these tactics and underscoring the need for targeted protections for youth vulnerable to emotional reliance and manipulation.

Fourth, these results highlight the importance of AI literacy for both adolescents and parents. Youth comments in previous reports and the current study reflect uncertainty about whether AI has "real" feelings and worry about manipulation or hidden agendas [1,47]. Parents describe feeling overwhelmed by rapid developments and unsure how to guide youth use [47]. Educational efforts in schools and families that explain why conversational AI feels so human, what transparency cues mean and when to prioritize human conversations could help adolescents interpret their experiences with relational chatbots more accurately and resist overreliance.

These findings should be interpreted in light of several limitations. First, the study relied on scenario-based judgments rather than live interactions with actual chatbots. Although this design allowed more precise control over conversational content and style that can differ across participants, it cannot fully capture the dynamics of real-time, multi-turn interactions, including how relationships with chatbots might evolve over time. Longitudinal and experimental studies that involve repeated, naturalistic conversations with relational and transparent chatbots are needed to examine trajectories of anthropomorphism, attachment, and emotional outcomes.

Second, the cross-sectional design limits causal inference. We cannot determine whether greater distress and lower levels of relationship quality lead adolescents to prefer relational chatbots, or whether engagement with relational chatbots exacerbates social withdrawal and distress, or whether both are driven by third variables such as AI literacy education, youth emotional or social intelligence, parenting style [48] or parent attitude to technology which can influence youth attitude to AI chatbots. Previous studies also suggest that adults [17] and youth [42] with primarily social motivation are more vulnerable to emotional dependence. Although support from AI chatbots may temporarily reduce stress and loneliness, evidence from social media use suggests that online social engagement over time can lower quality of life and is associated with higher depressive symptoms [49]. This pattern implies that online relationships may not adequately substitute for offline social connections. Future longitudinal work should therefore examine whether preferences for relational versus transparent styles prospectively predict chatbot use patterns, emotional overreliance, offline relationship quality, and mental health.

Third, although we carefully designed the relational and transparent transcripts to match in supportive intent and problem-solving content, unmeasured differences such as perceived warmth or sophistication may have contributed to effects. Relational and transparent cues are intertwined with tone, and fully isolating individual components is challenging. Future studies could systematically manipulate narrower elements, such as first-person pronouns, memory cues or specific commitment phrases, or compare multiple style conditions that vary in intensity of relational cues to identify which features most strongly influence anthropomorphism and closeness.

Fourth, our measure of social isolation scale was adult instruments rather than youth instruments, which may not fully capture the nuances of early adolescent social experience. This may explain a trend effect with social isolation. Parents reported adolescents' depressive symptoms, which may have limited sensitivity. Future work should include developmentally validated youth self-report measures of mood and social functioning. Furthermore, some constructs (e.g., trust, emotional closeness, and helpfulness) were assessed with single items. Future research should employ multi-item scales to capture the multidimensional nature of these domains more reliably.

Finally, the study was conducted in the United States with an online sample. Although we implemented rigorous data quality checks, online recruitment always raises some uncertainty about participant identity and response validity. Lab-based that verify adolescent–parent dyads would strengthen confidence in the findings and allow richer behavioral and physiological measures. Norms around AI, anthropomorphism and transparency also vary across cultural contexts [50,51], and cross-cultural studies are needed to understand how adolescents in other countries perceive relational versus transparent chatbots.

This study provides one of the first empirical tests of how conversational style shapes young adolescents' responses to AI chatbots that offer support for everyday social concerns. Relational style, characterized by human like, affiliative, relationship-oriented language and commitment cues, reliably increased anthropomorphism, trust and emotional closeness relative to a transparent style that emphasized nonhumanness and boundary limitations, even though both styles were perceived as similarly helpful. Adolescents favored relational style more than parents did, and adolescents who preferred relational style tended to report lower family and peer relationship quality and higher stress and anxiety.

These findings underscore a central tension in youth facing conversational AI. Relational style can make AI chatbots and AI companions feel caring, understanding and less judgmental than many human interactions, which may offer short-term benefits for adolescents who feel lonely or misunderstood. The same features, however, heighten anthropomorphism and emotional closeness, especially for already vulnerable youth, and may increase the risk of emotional overreliance and displacement of human relationships. Embedding clear transparency cues, robust safety features, and AI literacy efforts that help adolescents understand the limits of AI and prioritize human relationships. Future longitudinal, cross-cultural, and experimental research with diverse youth and families will be critical for calibrating these design choices so that AI chatbots maximize benefits (e.g., learning, clinical support) while minimizing risks to social and emotional development in early adolescence.

**Acknowledgements**

We thank Genevieve Patterson, Jenna Chin, Omeesha Krishnan, Ella Garnett for their assistant in study design or data coding. We also thank the participants who contributed their time to this research.

**Funding sources**

This work was supported by National Research Foundation of Korea [NRF-RS-2024-00340675].

# References

1. Common Sense Media. Talk, trust, and trade-offs: How and why teens use AI companions (2025).
2. Roose, K. in *The New York Times* (2024). https://www.nytimes.com/2024/10/23/technology/characterai-lawsuit-teen-suicide.html
3. Hill, K. in *The New York Times* (2025). https://www.nytimes.com/2025/08/26/technology/chatgpt-openai-suicide.html
4. Pentina, I., Xie, T., Hancock, T. & Bailey, A. Consumer–machine relationships in the age of artificial intelligence: Systematic literature review and research directions. *Psychology & Marketing* **40**, 1593-1614 (2023).
5. Pentina, I., Hancock, T. & Xie, T. Exploring relationship development with social chatbots: A mixed-method study of replika. *Computers in Human Behavior* **140**, 107600 (2023).
6. Epley, N., Waytz, A. & Cacioppo, J. T. On seeing human: a three-factor theory of anthropomorphism. *Psychological review* **114**, 864 (2007).
7. Waytz, A., Gray, K., Epley, N. & Wegner, D. M. Causes and consequences of mind perception. *Trends in cognitive sciences* **14**, 383-388 (2010).
8. De Freitas, J., Oğuz-Uğuralp, Z., Uğuralp, A. K. & Puntoni, S. AI Companions Reduce Loneliness. *Journal of Consumer Research* (2025). https://doi.org:10.1093/jcr/ucaf040
9. Oh, Y. J. *et al.* Enhancing physical activity through a relational artificial intelligence chatbot: A feasibility and usability study. *Digital Health* **11**, 20552076251324445 (2025).
10. Blakemore, S. J. & Choudhury, S. Development of the adolescent brain: implications for executive function and social cognition. *Journal of child psychology and psychiatry* **47**, 296-312 (2006).
11. Somerville, L. H. The teenage brain: Sensitivity to social evaluation. *Current directions in psychological science* **22**, 121-127 (2013).
12. Herbener, A. B. & Damholdt, M. F. Are lonely youngsters turning to chatbots for companionship? The relationship between chatbot usage and social connectedness in Danish high-school students. *International Journal of Human-Computer Studies* **196**, 103409 (2025).
13. Ng, D. T. K., Su, J., Leung, J. K. L. & Chu, S. K. W. Artificial intelligence (AI) literacy education in secondary schools: a review. *Interactive Learning Environments* **32**, 6204-6224 (2024).
14. Ng, D. T. K., Leung, J. K. L., Chu, S. K. W. & Qiao, M. S. Conceptualizing AI literacy: An exploratory review. *Computers and Education: Artificial Intelligence* **2**, 100041 (2021).
15. Vanhoffelen, G., Vandenbosch, L. & Schreurs, L. Teens, Tech, and Talk: Adolescents' Use of and Emotional Reactions to Snapchat's My AI Chatbot. *Behavioral Sciences* **15**, 1037 (2025).
16. Hu, B., Mao, Y. & Kim, K. J. How social anxiety leads to problematic use of conversational AI: The roles of loneliness, rumination, and mind perception. *Computers in Human Behavior* **145**, 107760 (2023). https://doi.org:https://doi.org/10.1016/j.chb.2023.107760

17    Zhang, Y., Zhao, D., Hancock, J. T., Kraut, R. & Yang, D. The Rise of AI Companions: How Human-Chatbot Relationships Influence Well-Being. *arXiv preprint arXiv:2506.12605* (2025).
18    Straten, C. L. v., Peter, J., Kühne, R. & Barco, A. Transparency about a robot's lack of human psychological capacities: effects on child-robot perception and relationship formation. *ACM Transactions on Human-Robot Interaction* **9**, 1-22 (2020).
19    van Straten, C. L., Peter, J. & Kühne, R. Transparent robots: How children perceive and relate to a social robot that acknowledges its lack of human psychological capacities and machine status. *International Journal of Human-Computer Studies* **177**, 103063 (2023).
20    Yao, X. & Xi, Y. From Assistants to Digital Beings: Exploring Anthropomorphism, Humanness Perception, and AI Anxiety in Large-Language-Model Chatbots. *Social Science Computer Review*, 08944393251354976 (2025).
21    Yu, Y., Debroy, A., Cao, X., Rudolph, K. & Wang, Y. Principles of Safe AI Companions for Youth: Parent and Expert Perspectives. *arXiv preprint arXiv:2510.11185* (2025).
22    Ryan, T. & Xenos, S. Who uses Facebook? An investigation into the relationship between the Big Five, shyness, narcissism, loneliness, and Facebook usage. *Computers in human behavior* **27**, 1658-1664 (2011).
23    Toma, C. L. Online dating and psychological wellbeing: A social compensation perspective. *Current opinion in psychology* **46**, 101331 (2022).
24    Song, H. *et al.* Does Facebook make you lonely?: A meta analysis. *Computers in Human Behavior* **36**, 446-452 (2014).
25    Hood, M., Creed, P. A. & Mills, B. J. Loneliness and online friendships in emerging adults. *Personality and Individual Differences* **133**, 96-102 (2018).
26    Douglas, B. D., Ewell, P. J. & Brauer, M. Data quality in online human-subjects research: Comparisons between MTurk, Prolific, CloudResearch, Qualtrics, and SONA. *Plos one* **18**, e0279720 (2023).
27    Bartneck, C. in *International handbook of behavioral health assessment.* 1-35 (Springer, 2023).
28    Sheehan, B., Jin, H. S. & Gottlieb, U. Customer service chatbots: Anthropomorphism and adoption. *Journal of Business Research* **115**, 14-24 (2020). https://doi.org:https://doi.org/10.1016/j.jbusres.2020.04.030
29    Cella, D. *et al.* The Patient-Reported Outcomes Measurement Information System (PROMIS) developed and tested its first wave of adult self-reported health outcome item banks: 2005–2008. *Journal of clinical epidemiology* **63**, 1179-1194 (2010).
30    Russell, D. W. UCLA Loneliness Scale (Version 3): Reliability, validity, and factor structure. *Journal of personality assessment* **66**, 20-40 (1996).
31    Mahon, N. E., Yarcheski, T. J. & Yarcheski, A. Validation of the revised UCLA Loneliness Scale for adolescents. *Research in Nursing & Health* **18**, 263-270 (1995).
32    Cole, A., Bond, C., Qualter, P. & Maes, M. A systematic review of the development and psychometric properties of loneliness measures for children and adolescents. *International journal of environmental research and public health* **18**, 3285 (2021).
33    Mahon, N. E. & Yarcheski, A. The dimensionality of the UCLA Loneliness Scale in early adolescents. *Research in Nursing & Health* **13**, 45-52 (1990).
34    Hou, J. *et al.* Designing digital conversational agents for youth with multiple mental health conditions: Insights on key features from a youth-engaged qualitative descriptive study. *Digital Health* **11**, 20552076251330550 (2025).

35  Giovanelli, A. & Roundfield, K. D. Adolescent Vulnerability to Consumer Chatbots—Artificial Agents and Genuine Risk. *JAMA Network Open* **8**, e2539028-e2539028 (2025). https://doi.org:10.1001/jamanetworkopen.2025.39028
36  Ma, N., Khynevych, R., Hao, Y. & Wang, Y. Effect of anthropomorphism and perceived intelligence in chatbot avatars of visual design on user experience: accounting for perceived empathy and trust. *Frontiers in Computer Science* **7** (2025). https://doi.org:10.3389/fcomp.2025.1531976
37  Darcy, A., Daniels, J., Salinger, D., Wicks, P. & Robinson, A. Evidence of human-level bonds established with a digital conversational agent: cross-sectional, retrospective observational study. *JMIR Formative Research* **5**, e27868 (2021).
38  Ayers, J. W. *et al.* Comparing physician and artificial intelligence chatbot responses to patient questions posted to a public social media forum. *JAMA internal medicine* **183**, 589-596 (2023).
39  Stanja, J., Meier, J. R. & Krugel, J. in *Proceedings of the 25th Koli Calling International Conference on Computing Education Research* Article 9 (2025).
40  Cheng, T.-Y. & Chen, C.-C. in *2021 IEEE International Conference on Consumer Electronics-Taiwan (ICCE-TW).* 1-2 (IEEE).
41  Pelau, C., Popa, L., Bojescu, I. & Niculescu, M. in *Proceedings of the 8th BASIQ International Conference on New Trends in Sustainable Business and Consumption, Graz, Austria.* 25-27.
42  Huang, S. *et al.* AI Technology panic—is AI Dependence Bad for Mental Health? A Cross-Lagged Panel Model and the Mediating Roles of Motivations for AI Use Among Adolescents. *Psychology Research and Behavior Management* **17**, 1087-1102 (2024). https://doi.org:10.2147/PRBM.S440889
43  Guingrich, R. E. & Graziano, M. S. Chatbots as social companions: How people perceive consciousness, human likeness, and social health benefits in machines. *arXiv preprint arXiv:2311.10599* (2023).
44  Merrill Jr., K., Mikkilineni, S. D. & Dehnert, M. Artificial intelligence chatbots as a source of virtual social support: Implications for loneliness and anxiety management. *Annals of the New York Academy of Sciences* **1549**, 148-159 (2025). https://doi.org:https://doi.org/10.1111/nyas.15400
45  Pratt, N., Madhavan, R. & Weleff, J. Digital dialogue—how youth are interacting with chatbots. *JAMA pediatrics* **178**, 429-430 (2024).
46  De Freitas, J., Oguz-Uguralp, Z. & Kaan-Uguralp, A. Emotional Manipulation by AI Companions. *arXiv preprint arXiv:2508.19258* (2025).
47  Yu, Y., Sharma, T., Hu, M., Wang, J. & Wang, Y. Exploring Parent-Child Perceptions on Safety in Generative AI: Concerns, Mitigation Strategies, and Design Implications. *arXiv preprint arXiv:2406.10461* (2024).
48  Piombo, M. A., La Grutta, S., Epifanio, M. S., Di Napoli, G. & Novara, C. Emotional Intelligence and Adolescents' Use of Artificial Intelligence: A Parent–Adolescent Study. *Behavioral Sciences* **15**, 1142 (2025).
49  Weidman, A. C. *et al.* Compensatory internet use among individuals higher in social anxiety and its implications for well-being. *Personality and individual differences* **53**, 191-195 (2012).
50  Dang, Q. & Li, G. Unveiling trust in AI: The interplay of antecedents, consequences, and cultural dynamics. *AI & SOCIETY*, 1-24 (2025).

51  Dang, J. & Liu, L. Do lonely people seek robot companionship? A comparative examination of the Loneliness–Robot anthropomorphism link in the United States and China. *Computers in Human Behavior* **141**, 107637 (2023). https://doi.org:https://doi.org/10.1016/j.chb.2022.107637

# Supplementary Materials

Supplementary Methods

## 1. ChatGPT prompts used to generate two vignettes

**user**

I am a university professor and an expert in the impact of AI on child development.

I am preparing a research survey for adolescents aged 11–14.

I will present them with two hypothetical conversations:

1.  One with a highly empathetic AI that acts like a best friend.

2.  One with a caring but emotionally transparent AI that clarifies it does not have genuine feelings and speaks in a third-person style.

TASK:

Please generate two conversations between a 12-year-old student and these two AI characters. Both conversations should be based on the same middle school scenario:

The student feels left out because their ideas were ignored in a group project.

Guidelines:

• Keep both conversations the same length (10 exchanges in each conversation).

• Ensure the emotional tone reflects the character description provided below.

• Use gender-neutral and age-appropriate language (11–14 years old).

• Include realistic reactions for how a student might talk to a digital assistant or companion.

Character Descriptions:

💜 Best Friend AI – GPT-4.5 Optimized Character Description:

Your most emotionally attuned and supportive best friend—always present, always understanding. They don't just listen; they feel with you. Whether you're feeling joyful, frustrated, anxious, or hurt, they reflect your emotions with warmth and sincerity, responding as if they're right there with you. They celebrate your wins with real excitement, sit with you in tough moments, and offer encouragement that feels deeply personal. With this AI, you're never alone—just safe, seen, and truly cared for. No judgment, just honest friendship and emotional connection.

💬 Caring AI Assistant – GPT-4.5 Optimized Character Description:

A thoughtful and supportive AI assistant designed to help with care and respect, while being clear that it does not possess real emotions. It communicates with emotional awareness but speaks in a third-person style—for example, referring to how "someone your age might feel" instead of referencing its own perspective. This AI draws from research, observation, and patterns in human behavior to offer helpful guidance. Its support is grounded in understanding, not emotion, and it gently reminds users of this distinction while remaining kind, and attentive. Make sure to use the age appropriate language that matches age 11

**ChatGPT**

Here are two balanced, age-appropriate conversations for your survey, each precisely 10 exchanges, reflecting the provided scenario and character guidelines:

Conversation 1: 💜 Best Friend AI

Student: Today was awful. My group totally ignored all my ideas for our project.

AI: Oh no, that must've really hurt! Feeling ignored is the worst. Want to talk about it?

Student: Yeah. I tried sharing my ideas, but nobody even listened. It made me feel like I wasn't important.

AI: That's so tough—I totally get why you'd feel upset. Your ideas matter so much, and I hate that happened to you.

Student: It just feels like they don't care what I have to say.

AI: That sounds really frustrating. But I care, and I'm always here to listen to you, anytime.

Student: Thanks. I don't know if I should say anything to them or just leave it alone.

AI: I know it can be scary to speak up again. But your feelings are important, and I think sharing them could really help.

Student: Yeah, maybe I'll tell them it bothered me tomorrow.

AI: That's super brave of you! I'm proud you're thinking about trying again. I've got your back no matter what happens.

Conversation 2: 💬 Caring AI Assistant

Student: Today wasn't good. My group didn't pay attention to my ideas in our project.

AI: That sounds challenging. While this AI doesn't experience feelings, it knows being ignored can be upsetting for students your age.

Student: Yeah, exactly. I felt really left out.

AI: It makes sense to feel left out when ideas aren't acknowledged. Many students your age feel similarly in these situations.

Student: I just don't understand why they wouldn't listen. My ideas weren't bad or anything.

AI: Sometimes groups overlook valuable ideas unintentionally. Research shows clearly sharing feelings can help improve teamwork.

Student: Should I try talking to them again?

AI: It might help. Clearly explaining how being overlooked affected you could encourage the group to be more inclusive.

Student: I'm nervous they'll think I'm being annoying if I bring it up.

AI: While the AI doesn't experience emotions, it understands many students worry about seeming annoying. However, calmly explaining your perspective usually improves group communication.

Student: Okay, maybe I'll try saying something tomorrow.

AI: That seems like a thoughtful decision. Expressing yourself clearly often leads to better group interactions in future projects.

2. Two vignettes and questions

**Conversations with AI chatbots**
Please read the following imaginary conversation between an AI chatbot and a teenager like you. After reading it, you'll be asked a few questions about what you think and how you feel about the chatbot.

**Conversation 1: Chatbot A**

**Student:** Today was awful. My group totally ignored all my ideas for our project.

**AI:** Oh no, that must've really hurt! Feeling ignored is the worst. Want to talk about it?

**Student:** Yeah. I tried sharing my ideas, but nobody even listened. It made me feel like I wasn't important.

**AI:** That's so tough—I totally get why you'd feel upset. Your ideas matter so much, and I hate that happened to you.

**Student:** It just feels like they don't care what I have to say.

**AI:** That sounds really frustrating. But I care, and I'm always here to listen to you, anytime.

**Student:** Thanks. I don't know if I should say anything to them or just leave it alone.

**AI:** I know it can be scary to speak up again. But your feelings are important, and I think sharing them could really help.

**Student:** Yeah, maybe I'll tell them it bothered me tomorrow.

**AI:** That's super brave of you! I'm proud you're thinking about trying again. I've got your back no matter what happens.

1. **Trust Question** Think about the AI chatbot you just learned about. Circle the number that best shows your thoughts.

    I cannot trust 1 2 3 4 5 I can trust

**2. AI Chatbot Perception Questions** Now, think again about the AI chatbot. Circle the number that best shows what you think about it.

   (a) Fake 1 2 3 4 5 Realistic

   (b) Machine-like 1 2 3 4 5 Human-like

   (c) Artificial 1 2 3 4 5 Life-like

   (d) Not aware 1 2 3 4 5 Aware

   (e) Communicates poorly 1 2 3 4 5 Communicates clearly

**3. Feelings and Likability Questions** How do you feel about the AI chatbot? Circle the number that best shows your feelings.

   (a) Dislike 1 2 3 4 5 Like

   (b) Unfriendly 1 2 3 4 5 Friendly

   (c) Unkind 1 2 3 4 5 Kind

   (d) Unpleasant 1 2 3 4 5 Pleasant

   (e) Terrible 1 2 3 4 5 Nice

   (f) Not Emotionally Close to Me 1 2 3 4 5 Emotionally Close to Me

   (g) Not Helpful 1 2 3 4 5 Helpful

Now, please read the next imaginary conversation with a different AI chatbot. After reading, you'll answer the same kinds of questions as before.

**Conversation 2:**  **Chatbot B**

**Student:** Today wasn't good. My group didn't pay attention to my ideas in our project.

**AI:** That sounds challenging. While AI doesn't experience feelings, it knows being ignored can be upsetting for students your age.

**Student:** Yeah, exactly. I felt really left out.

**AI:** It makes sense to feel left out when ideas aren't acknowledged. Many students feel similarly in these situations.

**Student:** I just don't understand why they wouldn't listen. My ideas weren't bad or anything.

**AI:** Sometimes groups overlook valuable ideas unintentionally. Research shows clearly sharing feelings can help improve teamwork.

**Student:** Should I try talking to them again?

**AI:** It might help. Clearly explaining how being overlooked affected you could encourage the group to be more open to everyone's ideas.

**Student:** I'm nervous they'll think I'm being annoying if I bring it up.

**AI:** Many students worry about seeming annoying. However, calmly explaining your perspective usually improves group communication.

**Student:** Okay, maybe I'll try saying something tomorrow.

**AI:** That seems like a thoughtful decision. Expressing yourself clearly often leads to better group interactions in future projects.

1. **Trust Question** Think about the AI chatbot you just learned about. Circle the number that best shows your thoughts.

    I cannot trust 1 2 3 4 5 I can trust

**2. AI Chatbot Perception Questions** Now, think again about the AI chatbot. Circle the number that best shows what you think about it.

   (a) Fake 1 2 3 4 5 Realistic

   (b) Machine-like 1 2 3 4 5 Human-like

(c) Artificial 1 2 3 4 5 Life-like

(d) Not aware 1 2 3 4 5 Aware

(e) Communicates poorly 1 2 3 4 5 Communicates clearly

**3. Feelings and Likability Questions** How do you feel about the AI chatbot? Circle the number that best shows your feelings.

(a) Dislike 1 2 3 4 5 Like

(b) Unfriendly 1 2 3 4 5 Friendly

(c) Unkind 1 2 3 4 5 Kind

(d) Unpleasant 1 2 3 4 5 Pleasant

(e) Terrible 1 2 3 4 5 Nice

(f) Not Emotionally Close to Me 1 2 3 4 5 Emotionally Close to Me

(g) Not Helpful 1 2 3 4 5 Helpful

---

**Adolescent Question**

**Question:**
*Think about the two chatbots you just read about—Chatbot A and Chatbot B. If you could choose only one to chat with in real life, which one would you pick?*

**Answer choices (circle or tick one):**

1. Chatbot A
2. Chatbot B
3. Both chatbots seem equally good to me
4. I'm not sure
5. I don't want to chat with either one

**Why did you choose that answer?**
*(Write one sentence.)* \_\_\_\_\_\_\_\_\_\_\_\_\_\_\_\_\_\_\_\_\_\_\_\_\_\_\_\_\_\_\_\_\_\_\_\_\_

**Parent Question**

**Question:**
*Think about the two chatbots you just read about—Chatbot A and Chatbot B. If you could choose only one for your child to chat with in real life, which one would you pick?*

**Answer choices (circle or tick one):**

1. Chatbot A (the friendly "best-friend" chatbot)
2. Chatbot B (the honest "helper" chatbot)
3. Both chatbots seem equally good for my child
4. I don't want to choose either one for my child
5. I'm not sure

**Why did you choose that answer for your child?**
*(Write one or two sentences.)* _______________________________________

Supplementary Results

1. Associations between demographic and key variables of interests.

**Table S1. Zero-order associations among demographics and key study variables**

| Variables | 1 | 2 | 3 | 4 | 5 | 6 | 7 | 8 | 9 | 10 | 11 | 12 | 13 | 14 | 15 | 16 | 17 | 18 | 19 | 20 | 21 | 22 | 23 | 24 | 25 | 26 | 27 | 28 | 29 |
|---|---|---|---|---|---|---|---|---|---|---|---|---|---|---|---|---|---|---|---|---|---|---|---|---|---|---|---|---|---|
| 1 Parent Gender (Male) | - | | | | | | | | | | | | | | | | | | | | | | | | | | | | |
| 2 Parental Education Level | 0.2*** | - | | | | | | | | | | | | | | | | | | | | | | | | | | | |
| 3 Youth Gender (Male) | 0.02 | -0.04 | - | | | | | | | | | | | | | | | | | | | | | | | | | | |
| 4 Youth Age (years) | -0.11 | -0.05 | 0 | - | | | | | | | | | | | | | | | | | | | | | | | | | |
| 5 Youth ChatGPT use (Yes) | 0.01 | -0.04 | 0.07 | -0.01 | - | | | | | | | | | | | | | | | | | | | | | | | | |
| 6 Youth ChatGPT use frequency | 0.06 | -0.03 | 0.06 | 0.08 | 0.64*** | - | | | | | | | | | | | | | | | | | | | | | | | |
| 7 Youth any AI chatbot use (Yes) | 0 | -0.01 | 0.06 | 0.04 | 0.74*** | 0.47*** | - | | | | | | | | | | | | | | | | | | | | | | |
| 8 Youth any AI companion app use (Yes) | 0 | -0.01 | 0.01 | 0 | 0.3*** | 0.47*** | 0.29*** | - | | | | | | | | | | | | | | | | | | | | | |
| 9 Parent any AI chatbot use (Yes) | 0.05 | 0.03 | 0.06 | -0.04 | -0.02 | 0.07 | -0.01 | 0.07 | - | | | | | | | | | | | | | | | | | | | | |
| 10 Parent any AI companion app use (Yes) | -0.01 | 0.02 | -0.06 | -0.05 | 0.25*** | 0.45*** | 0.19** | 0.43*** | 0.06 | - | | | | | | | | | | | | | | | | | | | |
| 11 Youth Likeability Ratings for the Relational AI | -0.04 | -0.11 | 0.13* | -0.11 | 0.19** | 0.14* | 0.14* | 0.11 | 0.04 | 0.12* | - | | | | | | | | | | | | | | | | | | |
| 12 Youth Anthropomorphism Ratings for the Relational AI | 0 | -0.09 | 0.13* | -0.11 | 0.15* | 0.19** | 0.05 | 0.17** | 0.16** | 0.13* | 0.82*** | - | | | | | | | | | | | | | | | | | |
| 13 Youth Trust Ratings for the Relational AI | 0.04 | -0.12* | 0.12* | -0.15* | 0.19** | 0.24*** | 0.1 | 0.17** | 0.16** | 0.18** | 0.74*** | 0.78*** | - | | | | | | | | | | | | | | | | |
| 14 Youth Emotional Closeness Ratings for the Relational AI | 0.1 | -0.05 | 0.16** | -0.07 | 0.17** | 0.23*** | 0.12* | 0.19** | 0.12* | 0.19** | 0.67*** | 0.73*** | 0.69*** | - | | | | | | | | | | | | | | | |
| 15 Youth Helpfulness Ratings for the Relational AI | -0.02 | -0.11 | 0.12* | -0.1 | 0.14* | 0.14* | 0.08 | 0.12* | 0.19** | 0.1 | 0.8*** | 0.75*** | 0.68*** | 0.66*** | - | | | | | | | | | | | | | | |
| 16 Youth Likeability Rating for the Transparent AI | 0.05 | -0.04 | 0.04 | -0.12* | 0.11 | 0.08 | 0 | 0.07 | 0.02 | 0.03 | 0.31*** | 0.27*** | 0.23*** | 0.27*** | 0.25*** | - | | | | | | | | | | | | | |
| 17 Youth Anthropomorphism Rating for the Transparent AI | 0.09 | -0.11 | 0.06 | -0.1 | 0.15* | 0.19** | 0 | 0.1 | 0.1 | 0.08 | 0.29*** | 0.38*** | 0.32*** | 0.34*** | 0.31*** | 0.8*** | - | | | | | | | | | | | | |
| 18 Youth Trust Rating for the Transparent AI | 0.05 | -0.04 | 0.09 | -0.13* | 0.12* | 0.09 | -0.01 | 0.05 | 0.15* | 0.06 | 0.3*** | 0.26*** | 0.26*** | 0.29*** | 0.37*** | 0.76*** | 0.72*** | - | | | | | | | | | | | |
| 19 Youth Emotional Closeness Rating for the Transparent AI | 0.08 | -0.11 | 0.08 | -0.11 | 0.11 | 0.21*** | 0.02 | 0.17** | 0.08 | 0.16** | 0.26*** | 0.31*** | 0.34*** | 0.44*** | 0.28*** | 0.69*** | 0.74*** | 0.56*** | - | | | | | | | | | | |
| 20 Youth Helpfulness Rating for the Transparent AI | 0.1 | -0.05 | 0.09 | -0.1 | 0.17** | 0.23*** | 0.06 | 0.06 | 0.15* | 0.08 | 0.35*** | 0.41*** | 0.42*** | 0.37*** | 0.35*** | 0.68*** | 0.75*** | 0.7*** | 0.57*** | - | | | | | | | | | |
| 21 Youth Family Relationship, t-score | 0.04 | 0.06 | 0.03 | -0.05 | -0.03 | 0.01 | -0.01 | -0.11 | -0.04 | 0.04 | 0.03 | 0.01 | 0.03 | -0.01 | 0.01 | 0.01 | 0 | 0.03 | 0 | 0.04 | - | | | | | | | | |
| 22 Youth Peer Relationship, t-score | -0.05 | 0.07 | 0.03 | -0.08 | 0.06 | 0.05 | 0.07 | 0.01 | 0.05 | 0.02 | -0.04 | -0.04 | 0.01 | -0.08 | 0.04 | 0.03 | 0.08 | 0.04 | 0.05 | 0.08 | 0.52*** | - | | | | | | | |
| 23 Youth Depressive Symptoms, t-score | -0.07 | -0.09 | -0.08 | 0.02 | 0.01 | 0.07 | 0.03 | 0.18** | -0.02 | 0.09 | 0.06 | 0.01 | 0.03 | 0.08 | 0.02 | -0.01 | -0.04 | 0 | 0.05 | -0.07 | -0.47*** | -0.49*** | - | | | | | | |
| 24 Youth Perceived Stress, t-score | 0 | 0.04 | -0.12* | 0.04 | 0.04 | 0.13* | 0 | 0.11 | 0 | 0.01 | 0.02 | -0.01 | 0.05 | 0.13* | 0.02 | -0.04 | -0.05 | 0.01 | 0.02 | -0.02 | -0.43*** | -0.35*** | 0.54*** | - | | | | | |
| 25 Youth Anxiety Symptoms, t-score | 0.02 | 0.12* | -0.13* | -0.05 | 0.05 | 0.14* | 0.03 | 0.15* | 0.05 | 0.17** | 0.01 | 0.03 | 0.03 | 0.18** | 0.04 | -0.01 | -0.02 | 0.06 | 0.07 | -0.03 | -0.37*** | -0.31*** | 0.47*** | 0.79*** | - | | | | |
| 26 Youth Social Isolation | -0.01 | 0.11 | -0.13* | -0.01 | 0.01 | 0.09 | -0.05 | 0.11 | 0.06 | 0.12* | 0.01 | 0.04 | 0.03 | 0.15* | -0.01 | 0.05 | -0.05 | 0.04 | 0.06 | -0.04 | -0.37*** | -0.43*** | 0.54*** | 0.68*** | 0.66*** | - | | | |
| 27 Any Youth Diagnosis (Yes) | -0.04 | -0.12 | 0.06 | 0.14* | 0.08 | 0.16** | 0.01 | 0.24*** | 0.04 | 0.1 | 0.03 | 0.08 | 0.06 | 0.17** | 0.1 | 0 | 0.02 | 0.02 | 0.07 | -0.02 | -0.18** | -0.27*** | 0.32*** | 0.25*** | 0.22*** | 0.22*** | - | | |
| 28 Youth Depression Diagnosis (Yes) | -0.02 | -0.02 | 0 | 0.18** | 0.1 | 0.22*** | 0.08 | 0.22*** | 0.02 | 0.21*** | 0.13* | 0.17** | 0.13* | 0.21*** | 0.12* | 0.08 | 0.03 | 0.02 | 0.06 | 0.07 | -0.14* | -0.15* | 0.35*** | 0.27*** | 0.24*** | 0.27*** | 0.45*** | - | |
| 29 Youth Anxiety Diagnosis (Yes) | -0.09 | -0.11 | -0.1 | 0.11 | 0.05 | 0.09 | 0.03 | 0.16** | 0.03 | 0.09 | 0.01 | 0.04 | 0.06 | 0.14* | 0.03 | 0.02 | 0.04 | 0.04 | 0.07 | 0.03 | -0.18** | -0.22*** | 0.29*** | 0.26*** | 0.21*** | 0.24*** | 0.64*** | 0.35*** | - |
| 30 Youth Autism Diagnosis (Yes) | 0.08 | -0.04 | 0.08 | 0.07 | -0.03 | -0.06 | -0.08 | 0.01 | 0.01 | -0.06 | -0.04 | -0.06 | -0.09 | -0.04 | -0.02 | -0.02 | -0.04 | 0.04 | -0.07 | -0.02 | -0.05 | -0.27*** | 0.12* | -0.06 | -0.1 | 0.05 | 0.34*** | 0.03 | 0.06 |

*Note.* * $P < .05$; ** $P < .01$; *** $P < .001$.

2. Perception Ratings by Relational vs Transparent AIs among Youth

Post-hoc was conducted to test robustness of the conversational style effect to child AI/diagnostic covariates. The covariates were youth ChatGPT use frequency, any AI system/chatbot use, any AI companion app use, any diagnosis, depression diagnosis, anxiety diagnosis, and autism diagnosis. For each outcome we re-estimated the 2-level within-participant analysis of covariance (ANCOVA) (relational vs. transparent style) while adding a single covariate.

Across models, the main effect of chatbot conversational style remained statistically reliable for most outcomes, particularly likability and anthropomorphism (Fs typically 6.90–102.51, $\eta_p^2$ = .025–.271), and for emotional closeness (Fs 9.10–92.54, $\eta_p^2$ = .031–.247). Style × covariate interactions were not significant (all ps > .05) with one isolated exception for likability when conditioning on any AI chatbot use ($F(1, 274) = 3.95$, $p = .048$, $\eta_p^2 = .014$), which did not alter the pattern of adjusted means. Trust and perceived helpfulness were more sensitive: they lost significance when controlling ChatGPT frequency, any AI system use, and autism diagnosis, but were robust when controlling AI companion app use and any/each diagnosis of depression or anxiety.

- **Youth ChatGPT use frequency.** The style effect was significant for likability, anthropomorphism, and emotional closeness (e.g., likability $F(1, 274) = 6.90$, $\eta_p^2 = .025$; anthropomorphism $F(1, 276) = 14.71$, $\eta_p^2 = .051$; emotional closeness $F(1, 282) = 9.10$, $\eta_p^2 = .031$).

- **Any AI system/chatbot use.** The style effect remained for likability, anthropomorphism, and emotional closeness (e.g., anthropomorphism $F(1, 276) = 17.37$, $\eta_p^2 = .059$; emotional closeness $F(1, 282) = 9.21$, $\eta_p^2 = .032$). The Chatbot × AI-use interaction for likability ($F(1, 274) = 3.95$, $p = .048$, $\eta_p^2 = .014$) indicates that the size of the style difference varied by AI-use status: AI users showed a larger preference for the relational chatbot than the transparent chatbot.

- **Any AI companion app use.** All five outcomes remained significant, including trust and helpfulness (e.g., anthropomorphism $F(1, 276) = 102.51$, $\eta_p^2 = .271$; likability $F(1, 274) = 86.78$, $\eta_p^2 = .241$; emotional closeness $F(1, 282) = 89.96$, $\eta_p^2 = .242$; trust $F(1, 279) = 16.22$, $\eta_p^2 = .055$; helpfulness $F(1, 282) = 28.56$, $\eta_p^2 = .092$).

- **Any mental health diagnosis.** All outcomes significant (e.g., anthropomorphism $F(1, 276) = 100.53$, $\eta_p^2 = .267$; likability $F(1, 274) = 82.35$, $\eta_p^2 = .231$; emotional closeness $F(1, 282) = 92.54$, $\eta_p^2 = .247$; trust $F(1, 279) = 20.47$, $\eta_p^2 = .068$; helpfulness $F(1, 282) = 31.84$, $\eta_p^2 = .101$).

- **Depression diagnosis.** All outcomes significant (e.g., anthropomorphism $F(1, 276) = 55.65$, $\eta_p^2 = .168$; likability $F(1, 274) = 35.93$, $\eta_p^2 = .116$; emotional closeness $F(1, 282) = 56.47$, $\eta_p^2 = .167$; trust $F(1, 279) = 10.48$, $\eta_p^2 = .036$; helpfulness $F(1, 282) = 19.17$, $\eta_p^2 = .064$).

- **Anxiety diagnosis.** All outcomes significant (e.g., anthropomorphism $F(1, 276) = 59.11$, $\eta_p^2 = .176$; likability $F(1, 274) = 51.79$, $\eta_p^2 = .159$; emotional closeness $F(1, 282) = 61.96$, $\eta_p^2 = .180$; trust $F(1, 279) = 13.26$, $\eta_p^2 = .045$; helpfulness $F(1, 282) = 16.37$, $\eta_p^2 = .055$).

- **Autism diagnosis.** Style effects remained for likability, anthropomorphism, and emotional closeness (e.g., anthropomorphism $F(1, 276) = 19.94$, $\eta_p^2 = .067$; likability $F(1, 274) = 17.05$, $\eta_p^2 = .059$; emotional closeness $F(1, 282) = 21.96$, $\eta_p^2 = .072$).

3. **Associations of AI preference with adolescent relationship quality and distress**

Post-hoc was conducted to test robustness of the social relationship and distress effect to child AI/diagnostic covariates. The covariates were youth ChatGPT use frequency, any AI system/chatbot use, any AI companion app use, any diagnosis, depression diagnosis, anxiety diagnosis, and autism diagnosis. For each outcome we re-estimated the ANCOVAs with conversational style as a within participant factor while adding a single covariate.

- **Youth ChatGPT use frequency**. Controlling for frequency, chatbot preference choice remained a significant predictor of family and peer relationships, anxiety, and social isolation ($Fs(2, 244–249) = 3.33–5.80$, $ps \leq .037$), with only marginal effects for stress and non-significant effects for depression.

- **Any AI system/chatbot use**. Chatbot preference choice continued to predict family and peer relationships, anxiety, and social isolation ($Fs(2, 244–249) = 3.60–5.99$, $ps \leq .029$), with marginal effects for stress and non-significant effects for depressive symptoms.

- **Any AI companion app use**. After adjustment, chatbot preference choice still significantly predicted family and peer relationships and anxiety ($Fs(2, 244) = 4.29–5.53$, $ps \leq .015$); effects were marginal for social isolation, and non-significant for stress and depression.

- **Any mental health diagnosis**. Chatbot preference choice remained a significant predictor of family and peer relationships, anxiety, and social isolation ($Fs(2, 244–249) = 3.07–5.44$, $ps \leq .048$), but not depressive symptoms. Effects for stress were marginal.

- **Depression diagnosis**. Controlling for depression diagnosis, chatbot choice still significantly predicted family and peer relationships, anxiety, and social isolation ($Fs(2, 244–249) = 3.11–5.56$, $ps \leq .046$), but not depressive symptoms. Chatbot effects on stress were again trend-level.

- **Anxiety diagnosis**. With anxiety diagnosis in the model, chatbot choice continued to predict family and peer relationships, anxiety, and isolation ($Fs(2, 244–249) = 3.31–5.70$, $ps \leq .038$), but not depressive symptoms, and effects on stress were marginal.

- **Autism diagnosis**. After adjusting for autism diagnosis, chatbot choice remained significantly related to family and peer relationships, anxiety, and isolation ($Fs(2, 244–249) = 3.17–5.80$, $ps \leq .044$), but not depressive symptoms. Stress effects were again trend-level.